# Reptilian Skin as a Biomimetic Analogue for Design of Deterministic Tribo-Surfaces

H. A. Abdel-Aal [1,*], M. El Mansori[1], I.C. Gebeshuber [2,3,4,5]

[1] Arts et Métier ParisTech, Rue Saint Dominique BP 508, 51006 Chalons-en-Champagne, France

[2] Institute of Microengineering and Nanoelectronics (IMEN), Universiti Kebangsaan Malaysia, 43600 UKM, Bangi, Selangor, Malaysia

[3] TU BIONIK Center of Excellence for Biomimetics, Vienna University of Technology, Getreidemarkt 9/134, 1060 Wien, Austria

[4] Institute of Applied Physics, Vienna University of Technology, Wiedner Hauptstrasse 8-10/134, 1040 Wien, Austria

[5] AC²T Austrian Center of Competence for Tribology, Viktor Kaplan-Straße 2, 2700 Wiener Neustadt, Austria

[*] Contact author. Email: *Hisham.abdel-aal@chalons.ensam.fr*

**Abstract**

A major concern in designing tribo-systems is to minimize friction, save energy, and to reduce wear. Satisfying these requirements depends on the integrity of the rubbing surface and its suitability to sliding conditions. As such, designers currently focus on constructing surfaces that are an integral part of the function of the tribo-system. Inspirations for such constructs come from studying natural systems and from implementing natural design rules. One species that may serve as an analogue for design is the Ball python. This is because such a creature while depending on legless locomotion when sliding against various surfaces, many of which are deemed tribologically hostile, doesn't sustain much damage. Resistance to damage in this case originates from surface design features. As such, studying these features and how do they contribute to the control of friction and wear is very attractive for design purposes. In this work we apply a multi scale surface characterization approach to study surface design features of the Python regius that are beneficial to design high quality lubricating surfaces (such as those obtained through plateau honing). To this end, we studied topographical features by SEM and through White Light Interferrometery (WLI). We further probe the roughness of the surface on multi scale and as a function of location within the body. The results are used to draw a comparison to metrological features of commercial cylinder liners obtained by plateau honing.

**Introduction**

One of the current technical challenges facing engineers is to reduce global consumption of fossil fuels. At the core of the strategy devised to achieve such a goal is to reduce energy wasted by Internal Combustion engines (ICEs). ICEs typically operate with a thermal efficiency, ratio of output energy to input energy between 50% and 60% [1]. It is estimated [2] that roughly 15% of the energy input to a passenger vehicle is consumed by friction. In the United States alone there are more than two hundred million motor vehicles powered by ICEs. Many of such engines produce power in the order of one hundred kilowatts ($10^2$ KW). So that, friction-induced power losses translate into a staggering amount of wasted crude oil (estimated to be in excess of one million barrels a day in the United States alone).

Power losses due to friction represent a significant fraction of the overall power produced by an ICE. Piston-cylinder friction contributes a fare share of such lost power. Consequently, minimization of the friction losses due to the operation of these components is a major concern for engine designers. Typically, ICEs are lubricated to minimize friction between moving parts. Lubrication, however, doesn't eliminate friction and upon its' failure may cause significant damage to the engine. Further, due to the pressures acting on the piston and the cylinder, hydrodynamic stresses will develop in the oil films. These may contribute to frictional losses through fluid friction resulting from viscosity effects. An engine cylinder will contain very hot gases resulting from ignition of fuel. The high temperatures encountered may lead to failure of the lubrication layer and hence the occurrence of metal-to-metal contact between cylinder and piston.

Several factors affect the energy consumed in friction between solids. These may be broadly classified as intrinsic factors which pertain to the physical and chemical properties of the rubbing materials, and extrinsic factors that relate to applied loads, ambient operation temperatures, rubbing speeds, and topography of the rubbing surfaces, etc.,. Within the extrinsic factor group, surface topography crucially influences the integrity of the rubbing pair. This is because the surface of a rubbing interface manifests the changes that a solid undergoes during the manufacturing stages. It also reflects the response of the contacting solids to other extrinsic factors. More importantly, the topography of the rubbing interface affects the quality of lubrication. Therefore, constructing a surface of a predetermined topography, which yields a predictable response, and in the meantime self-adapts to changes in sliding conditions, can reduce frictional losses. Such surfaces, termed here as "deterministic surfaces" manifest an emerging trend. Many modern engines involve artificial textures that ornament the inner surface of cylinder bores (cylinder liner). There are several methods used to emboss the artificial textures (e.g., multistep honing, helical honing, controlled thin layer deposition and laser texturing) [3-12]. To date, however, there is no agreement on the optimal topology that such surfaces should acquire. Furthermore, a systematic methodology that, if applied, may generate deterministic surface designs, which meet particular functional requirements of a given engine, is virtually non-existent. Reasons for such a condition are rooted in the analytical philosophy that inspires design generation in Man Engineered Systems (MES). Such a philosophy fundamentally differs from the design paradigm that governs generation of design in natural systems.

Generation of design in natural systems (geometry, pattern, form, and texture) is a holistic phenomenon that synchronizes all design constituents toward an overall optimized performance envelope. Such an approach yields deterministic design outputs that while conceptually simple are of optimized energy expenditure footprint. Natural engineering, thus, seeks trans-disciplinary technically viable alternatives which, given functional constraints, require minimum effort to construct and economizes effort while functioning. In nature, there are many examples of designs and technical solutions that economize the effort needed for operation and minimize damage profiles [13]. An analogous design paradigm, within the *MES*, domain, has not matured as of yet. In addition, the process of texturing is a multistep operation of a high degree of functional complexity. This presents considerable challenges to surface designers who conceptually conform to conventional paradigms that do not acknowledge functional complexity to start with.

Confining the discussion to texturing by honing, we may identify two sets of interrelated parameter groups within textures of cylinder bores: product functionality factors and a so-called cylinder liner features. Oil consumption, for example, falls under the first parameter group. Each constituent of a parameter group, in turn, relates to a feature factor of the honing process itself. In our example, oil consumption relates to the honing tool angle of application, any chatter marks, residual grooves and cracks present within the virgin surface. The running-in performance of the cylinder-piston assembly originates from the mechanics of the intermediate step of the texturing process (the so-called plateaux forming stage) and whether foreign bodies are present in the material of the surface to start with. In all, there are four hundred parameters involved in artificial texturing of cylinder bores by honing. Consideration of all parameters is considerably difficult if approached by conventional means. To this end, it is necessary to devise a new alternative that permits by passing detailed consideration of the complexity of the process. Additionally, this alternative approach should allow harnessing the complexities of texturing to generate a deterministic design output in the natural sense. That is, a surface that self adapts to changing external contact conditions, and is able to optimize its' tribological response based on intrinsic features rather than external modifiers. To devise the alternative approach one has to define precisely the function of the desired surface. This, in principal, determines the objective of the design process and assists envisioning the final output.

The ultimate function of cylinder bore texturing is to reduce friction between the piston and cylinder walls while reducing the friction stresses on the oil film used in lubrication. This goal is mainly achieved through optimized topographical features of the rubbing surfaces. The mechanistic principle that aids in friction reduction is minimization of contact between cylinder and piston. Once minimization takes place, the overall frictional forces reduce and damage reduces as well. However, due to sliding wear will take place. This will alter the contact conditions because of the induced changes in topography. The required surface has to maintain optimal sliding performance despite topography modifications. That is the texturing has to self-adapt to induced changes in sliding conditions. Through this adaptation, the surface will maintain optimal performance within an envelope of conditions rather than achieving peak performance at a point. Such a requirement is a characteristic of natural systems. As such, inspiration for the envisioned texture has to originate from a natural system analogue. Squamate Reptiles (SR), are major inspirations in that context. They present diverse examples where surface structuring, and modifications through submicron and nano-scale features, achieves frictional regulation manifested in reduction of adhesion [14], abrasion resistance [15], and frictional anisotropy [16].

Squamata comprises two large clades, Iguania (about 1,230 known species) and Scleroglossa (about 6,000 known species), 3,100 of which are traditionally referred to as "lizards," and the remaining 2,900 species as "snakes" [17]. Squamates have a wide distribution all over the planet. They are found almost everywhere on earth except where factors that limit their survivability are present (e.g., higher altitudes where very cold temperatures are predominant year round). Their ecological diversity, and thereby their diverse habitat, presents a broad range of tribological environments many of which are hostile in terms of sliding and contact conditions. Such a situation requires specific tribological response that manifests itself in functional practices and surface design features. As such, Squamates offer a great resource that can be mined for viable surface design inspirations. Many studies describe appearance and structure of skin in Squamata [18-21]. Studies also describe the geometrical features and the evolution of functional adaptability of many species [22]. Tribological performance of the snake clade was also a subject of many studies in Biology, Herpetology, and engineering [23]. Researchers have studied the mechanical behavior of snakeskin [24, 25]. Design of bio-inspired robots inspired several investigations of snakes to understand the mechanisms responsible for regulating legless locomotion [26]. The design of lightweight high-resolution infrared sensors prompted the study of light detection in vipers [27]. Hazel et al [20] probed some of the nano-scale design features of three snake species. The authors documented the asymmetric features of the skin ornamentation to which both authors attributed frictional anisotropy. Shafiei and Alpas [28, 29] reported that snakeskin replicas provide anisotropic tribological properties that minimize frictional interaction. Such an effect, these authors argue, stems from the asymmetric shape of the protrusions at the ridges of the skin's scales. Functional adaptation and morphing of snake ornamentation was also subject to several studies. The results attribute adaptation to the presence of submicron-and nano-sized fibril structures acting to modify friction and adhesion during locomotion.

Muscular activity, sequence of contraction and relaxation of appropriate muscle groups, is the source of motion in snakes. Transfer of motion between the body of the snake and the substrate depends on generation of sufficient tractions. Thus, transmission of locomotion tractions and accommodation of motion, takes place through the skin. The skin of the snake assumes the role of motion transfer and accommodation of energy consumed during the initiation of motion. The number, type and sequence of muscular groups responsible for the initiation of motion, and thus employed in propulsion, will vary according to the particular mode of motion initiated. It will also depend on the habitat and the surrounding environment. This also will affect the effort invested in initiation of motion and thereby affects the function of the different parts of the skin and the amount of accommodated energy. Therefore, in general, different parts of the skin will have different functional requirements. Moreover, the life habits of the particular species, e.g., defense, hunting, and swallowing) will require different deterministic functions of the different parts of the skin.

In a recent study, Abdel-aal et al [30], used a Multi-Scale Surface Analysis Technique (MSST) to decode the design features of shed skin obtained from Python regius (Ball Python). Results pointed at the importance of localized surface design where surface topology, texture and form vary in accordance with specific functional requirements of different zones along the body. Moreover, the metrological features of the shed skin revealed a multi scale nature of the topographical makeup of the surface, with each scale targeted at optimal function at a particular scale of contact. This customization strongly relates to optimized performance in terms of minimized surface damage and possible economy of energy consumed in combating frictional tractions during locomotion.

The principal idea of artificial texturing is to enhance the sliding performance of a cylinder by imposing a geometrical pattern on the external layer of the liner. That is to create a predetermined geometrical ornamentation on the contacting surface of the cylinder liner. Texturing of the surface results in the creation of a protrusion above the surface, *plateau*, and a channel between any two protrusions, *groove*. Together, the plateau and the groove help reduce the destructive effects of sliding between piston and cylinder. The grooves retain lubrication oil during piston sliding and thereby they can replenish the lubrication film in subsequent sliding cycles. The plateau, meanwhile, provides raised cushions (islands), that the piston will contact upon sliding, so that reduction of the total contact area takes place. A system of such nature is multi-scale by default in the sense that the basic metrological characteristics of a honed surface will have values that depend on the scale of observation. It is this fact that renders the design of a honed surface inspired by the metrological features of a Python surface technically attractive. Multi-Scale features of conventional textured surfaces achieve optimal performance within a narrow domain of tribo-conditions, whereas Multi Scale features of the Python are optimal for a considerably broader range of rubbing situations. As such, if we understand topology and construction of the python skin, and their relation to sliding performance of the reptile, then we may apply the same design principles to devise an optimal textured surface fit for a broad envelope of operation conditions.

This chapter details an effort to deduce design rules that allow the construction of bio-inspired deterministic surfaces. We draw analogy between the topographical features of the Python regius shed skin and those of typical honed surfaces. The scope of the presentation is limited to the metrological analysis of the topographical features rather than delving into the consequence of the surface geometry on frictional response. The presentation proceeds as follows: In the first part, we provide background information about the species under study, its biological features, skin morphology, and essential features of the skin shedding process. The second section of the manuscript details microscopy observations of the shed skin. Further, we report within the second section, on the metrological aspects of the shed skin topography. The final part of the manuscript presents a comparative analysis between the topographical features of the shed skin and those of typical honed surfaces. Here we identify the differences between natural surface texturing and that implemented through conventional approaches.

## 2. Background

### 2.1 The python species

Python regius, figure 1, is a constrictor type non-venomous snake species that typically inhabits Africa. An adult snake may reach 90-120 cm long. Females tend to be rather larger than males. They may reach 120-150 cm long. The build of the snake is non-uniform as the head-neck region, as well as that of the tail, is thinner than the trunk -region. The trunk meanwhile is the region of the body where most of the snake body mass is concentrated. It is thicker than other parts. The tail section is rather conical in shape (figure 1-a). The overall cross section of the body is more elliptical than circular and the perimeter of the cross section is not uniform along the body. The ventral (stomach) part of the body is typically cream or extremely light yellow in color with occasional black spots scattered within (figures 1-b and 1-d). Skin of the python contains blotches imposed upon an otherwise black background. Shape of the blotches is non-uniform and their colors are dark and light brown (figure 1-c). Form of the body reflects on the contact behavior while sliding. The head will establish minimal contact and so will the tail. The trunk section, however, will be the region where most of the contact interaction takes place. In addition, due to the non-uniform shape most of the weight of the animal will be concentrated in the trunk section. Consequently, the trunk will constitute the section where most of the volume of the snake establishes contact with the substrate upon sliding. This implies that the skin within the trunk section will accommodate generation of tractions due to muscle contractions and also will exhibit the effect of the frictional tractions that resist locomotion.

Scales of various shapes and sizes cover the skin. The scales form by differentiation of the underlying skin (the epidermis). The number, arrangement, size and shape of the scales vary greatly from one species to another, but for each species, they are genetically fixed. A snake within a particular species will hatch with a fixed number of scales. The number of scales doesn't increase nor reduce as the snake matures. Scales, however, grow larger to accommodate growth and may change shape with each molt. Therefore, the patterns of scalation provide simple and accessible recognition characteristics for the taxonomic classification of snake species. The simple body scales overlap one another slightly. Further, a thin inter-scale layer of skin tissue continuously links them. Snakes periodically molt their scaly skins and acquire new ones.

### 2.2 Structure of snake skin

The skin of a snake is a complicated structure. Normally it contains two generations of skin. This is illustrated in figure 2. The figure shows an "*outer generation layer*" and an "*inner generation layer*". The first represents the layer of skin which is about to be shed. The second represents the layer about to replace the shed skin. Both layers are of the same compositional structure. Like that of many vertebrates, has two principal layers: the *dermis* which is the deeper layer of connective tissue with a rich supply of blood vessels and nerves; and the *epidermis,* which in reptiles consists of up to seven sub-layers or "*strata*" of closely packed cells, forming the outer protective coating of the body [31]. The "epidermis" has no blood supply, but its' inner most living cells obtain their nourishment by the diffusion of substances to and from the capillaries at the surface of the "dermis" directly beneath them.

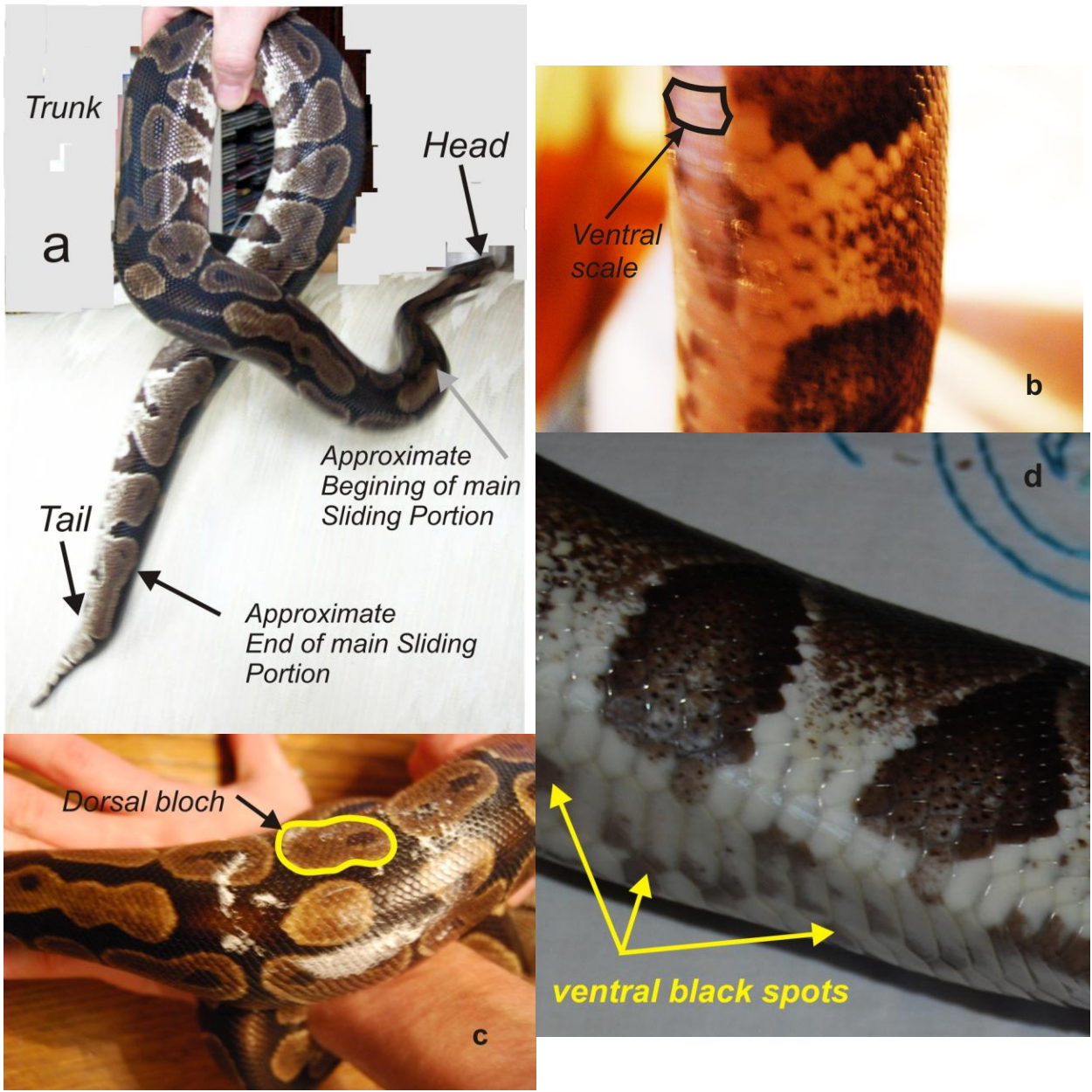

*Figure 1: General appearance of the Python regius. a, major regions of the body classified from a locomotion and contact mechanics point of view. Note the non-uniform distribution of the body cross sectional perimeter and that the head and tail sections are slender compared to the trunk. b, ventral side of the snake which is cream in color. C, d, dorsal blotches and ventral black spots*

The epidermis is the layer that directly contacts the surroundings. There are seven epidermal layers as shown in figure 2. The *"stratum germinativum"* is the deepest layer lining. It contains cells that have the capacity for rapid cell division. Six layers form each "*epidermal generation*" (the old and the new skin layers).

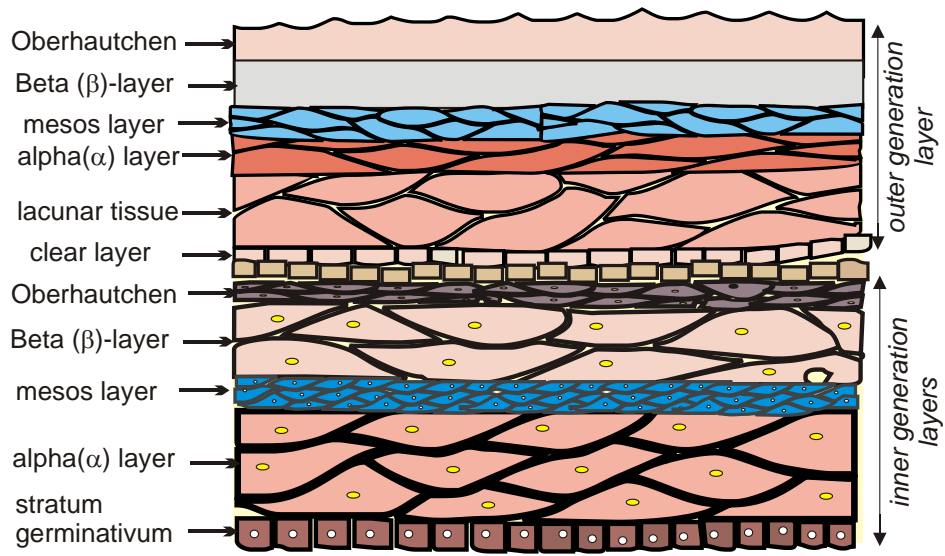

*Figure 2: Schematic illustartion of a generalized epidermis of a squamate reptile. Two layers are shown: the outer generation layer which represents the skin layer about to be shed and the inner generation layer that represents replacement of the shed layer.*

These are the clear layer and the *lacunar layer*, which matures in the old skin layer as the new skin is growing beneath. The alpha *(α)–layer*, the *mesos layer* and the *beta (β)-layer*, these layers consist of cells which are becoming keratinized with the production of two types of keratin (α and β keratin). These cells are thus being transformed into a hard protective layer. The final layer is the *"oberhautchen"* which constitutes the strongest outermost layer. It consists of a highly cornified (keratinized) surface, which is covered by a microscopically fine pattern of keels that is a species characteristic. The basale membrane separates the epidermis from the true inner skin (cutis or corium). The cutis consists of a more open and a more solid layer of connective tissue in which there is collagen and elastic fibers.

### 2.3 Skin shedding

In most mammals, the structure of the epidermis is less complex and the outermost dead skin cells are constantly flaking off; this protective layer constantly replenish from below. The deepest layer of cells, the "*stratum germinativum*", is constantly dividing and multiplying, and so the layers are on the move outwards [32]. In reptiles, however, this cell division, in the "*stratum germinativum*", only occurs periodically [33], and when it does all the layers above it, in the area where the cell division occurs, are replaced entirely. That is, the reptile, grows a second skin underneath the old skin, and then "sheds" the old one. About two weeks before the reptile sheds its skin, the cells in the *stratum germinativum* begin active growth and a second set of layers form slowly underneath the old ones. At the end of this time, the reptile effectively has double skin. Following such a process, the cells in the lowest layers of the old skin, the clear and the lacunar layers, and the Oberhautchen layer of the skin below undergo a final maturation and a so called "shedding complex forms". Fluid is exuded and forms a thin liquid layer between them. This gap between the two skins gives a milky appearance to a shedding reptile. Enzymes, in this fluid, break down the connections between the two layers.

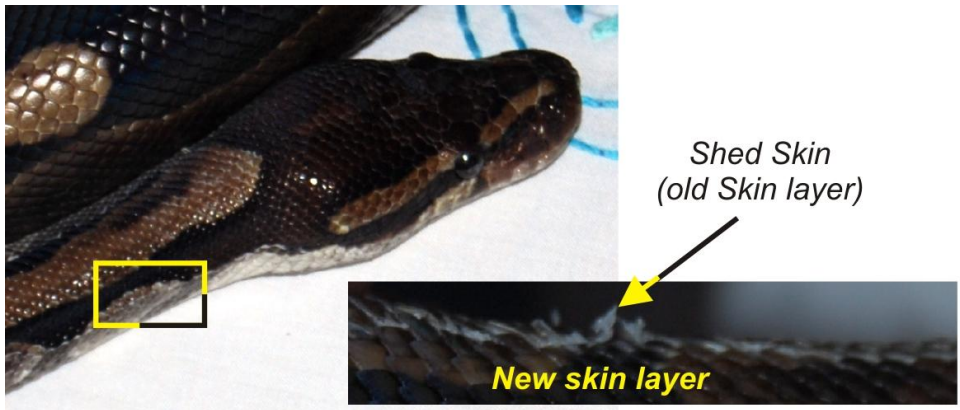

*Figure 3: Photograph of a Python regius during shedding. Two skin layers may be identified: the old (shed) layer almost milky in color, and the new replacement which is shiny. Note the flakey appearance of the skin within the magnified zone.*

The old skin lifts and the reptile actively removes it. This process is shown in figure 3. The figure depicts a live species undergoing skin shedding. Note the presence of two skin layers: the newly generated layer (shinny skin in figure) and the layer that is being shed (milky or opaque appearance).

### 3. Observation of shed skin
### 3.1 Initial observations

Initial observations of the scale structure were performed using photography of a live snake and optical microscopy. All observations took place without treatment of the skin. Figure 4 (a through e) details the surface structure of the live snake. To facilitate the analysis we defined two virtual axes on the body of the reptile. The first is designated as the Major Longitudinal Body Axis (MLBA). This line coincides with forward motion of the snake. It is a line centered on the hide of the beast and extends from the tip of the mouth to the tip of the tail (on the: dorsal and ventral sides). The second axis is a moving Transverse Axis (TRANS). This line coincides with slithering motion. In particular it coincides with lateral displacements of the body. An illustration of the axes on the head of the animal, on both sides, is provided in figures 4 a and b.

Figure 4 (a-c) details the skin geometry in the head region from the inner side (sliding side). A general view of the underside of the head is given in figure 4-a. A general view of the outer side of the head is provided as figure 4-b. Close up of the head-jaw area from the ventral side is provided as figure 4-c. The photographs reveal that polygons constitute the geometrical building block of the surface. This polygon has eight sides, octagon, in the general area of the mouth (represented by the letter *O* in the figure). Past a line that joins the eyes, line AA, the pattern of the skin changes to hexagonal. The hexagonal structure is also dominant within the outer (upper side) of the head as shown in figure 4-b. The size of the hexagonal cells differs from that of the octagonal cells. The size of octagons in the mouth region is not uniform. However, compared to the hexagonal patterns within the throat region, the area of the unit octagon is apparently greater than that of the hexagonal unit. Hexagonal cells on the other hand are of uniform shape and size and seem to be of uniform density per unit area. Within the mouth zone (above the line AA) the aspect ratio of the octagonal unit cell is, qualitatively, uniform. The major axis of the polygon, moreover, appears to point in the same direction of the snake. The uniform distribution of hexagonal cells is likely to aid the compliance of the surface and increase its flexibility. Such a hexagonal pattern is noted to be the most efficient way to pack the largest number of similar objects in a minimum space [34]. Figure 4-e depicts the surface geometry in the ventral side (general region of the belly). A hexagonal pattern constitutes the basic building block of the skin. The size of the hexagons differs around the circumference of the body. Large cells are particular to the main sliding area whereas cells of smaller size are particular to the back and the sides. The aspect ratio of the cells is variable. Of interest is the orientation of the major axis of the skin unit cells with respect to the MLBA. In the head-throat region the major axis of the cells is oriented parallel to the body major axis whereas in the ventral scales the major cell axis is perpendicular. Varenberg and Gorb [35], based on experiments on the hexagonal structures found on tarsal attachment pads of the bush cricket (*tettigonia viridissima*) suggest that variation in the aspect ratio of hexagonal structures may alter the friction force of elastomers by at least a factor of two. Additionally, we propose that the perpendicular orientation of the cells, with respect to the major axis of the snake, within the main sliding region aids in shifting the weight, and hence the contact angle and area of the snake upon sliding. Note that since the body of the snake is of cylindrical shape, the highest curvature of the skin will be oriented along the major cell axis. As such, upon sliding, the area of contact, and therefore the total tractions, will depend on the direction of motion (higher sideways and minimal forward). The orientation of the hexagon axis renders the friction forces anisotropic. Such an observation is consistent with the findings of Zhang et al [36] who studied the frictional mechanism and anisotropy of Burmese python's ventral scales. They reported that the friction coefficient of the ventral scale had closely relationship with moving direction. The frictional coefficient for backward and lateral motion was one third higher than that in forward motion.

**3.2    Optical Microscopy observations**

In snakes the *Oberhautchen* layer is in direct contact with the environment and possesses a fine surface structure called micro-ornamentation [37]. Earlier authors [37-39] described details of the micro-ornamentation. Initial observations on the structure of the scales were performed using optical microscopy without any treatment of the skin. Figure 5 depicts the structure of the scales at two positions within the skin in a region close to the waist of the snake. The first was from the back (dorsal scale) whereas the second position represented the stomach of the snake (ventral). Note that although the general form of the cells is quite similar for both positions the size of a unit cell within the skin is quite different in both cases. In particular the cell is wider for the ventral positions. Each cell (scale) is also composed of a boundary and a membrane like structure. Note also the overlapping geometry of the skin and the scales (the so called scale and hinge structure). The skin from the inner surface hinges back and forms a free area, which overlaps the base of the next scale, which emerges below this scale figure 5-b.

Figure 6 (a-c) provides details of the ventral scales. It is noted that the edges of the ventral scales are not straight. Rather, they are curved in the head tail plane with the curvature concave toward direction of the head (the arrow labeled H in figure 6-a). The overlapping arrangement of the scales and the existence of the elastic connecting tissue are shown in figure 6-b. The figure indicates that the curvature of the leading edge of any one scale (the edge toward the head side) is larger than of the trailing edge (figure 6-b). A close-up of the hinge region between two cells is depicted in figure 6-c. The hinge is made of the connecting elastic tissue. Note the crisscrossed pattern that distinguishes the elastic tissue area. Note also the size and pattern of the elastic connector tissue (termed as the membrane in figure 6-c) and how it surrounds individual cells.

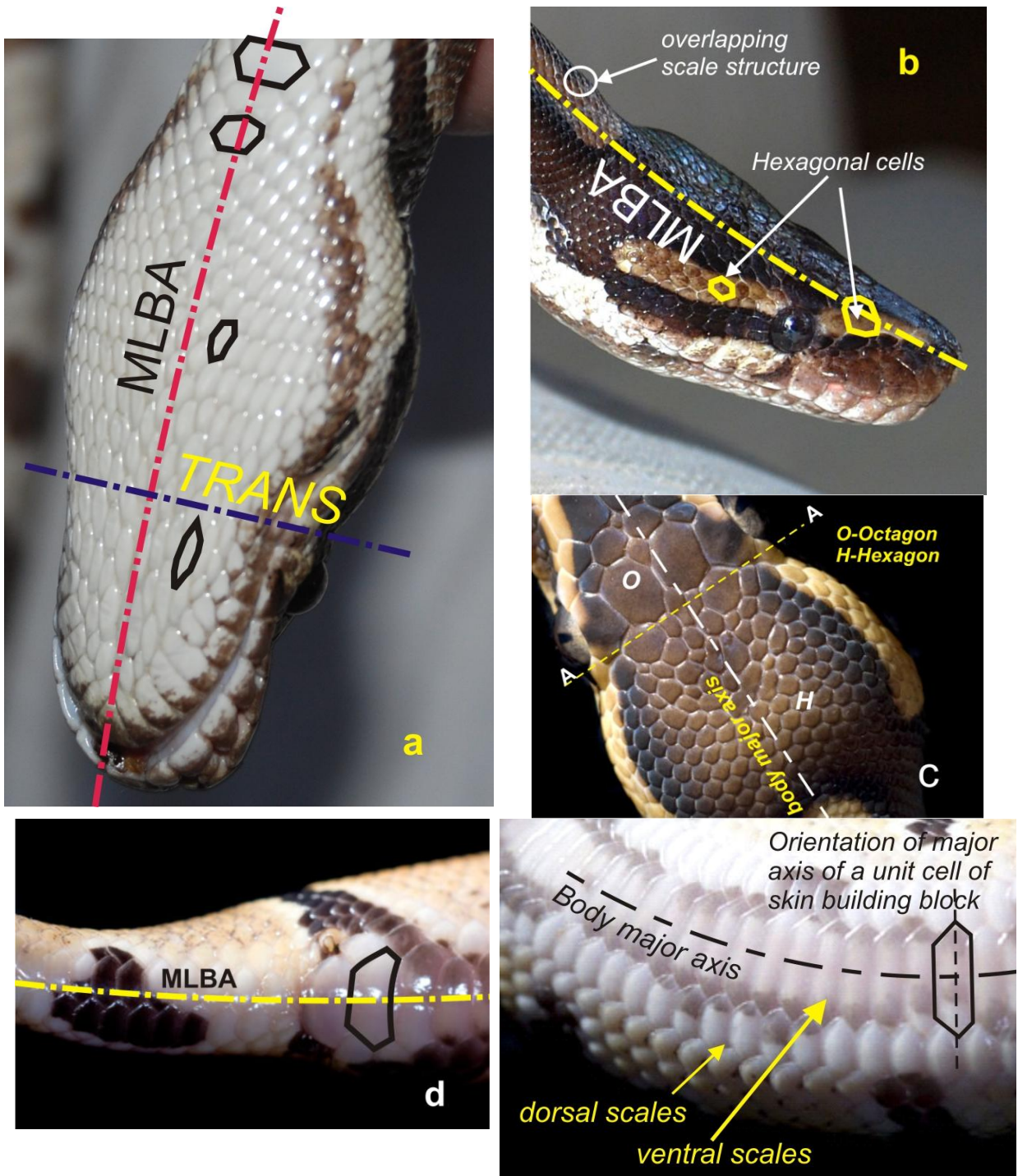

*Figure 4: Details of scale structure at several positions on the live snake. Figure 4-a: provides a generalized view of the throat-jaw region from the sliding side. Figure 4-b provides a general view of the outer side of the head. Figure 4-c presents a close-up of the jaw throat area, here the different polygon-based structures of the unit building blocks of the skin are identified. Hexagonal cells are designated H and Octagonal cells are labeled O. Note the difference in size between octagonal and hexagonal cells. Figure 4-d depicts skin structure within the tail section and figure 4-e details the structure within the ventral side of the trunk. Notice that the hexagonal cells have high aspect ratio and that the longest diagonal is oriented perpendicular to the MLBA*

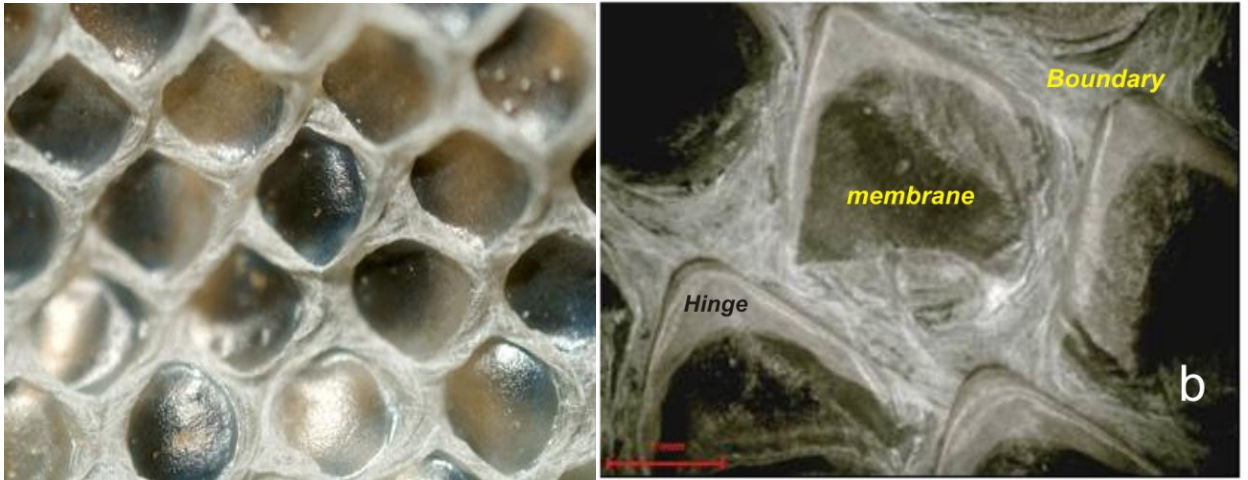

*Figure 5. The structure of the scales on the inside of the shed skin at a region close to the mid section of the species at two orientations: back (dorsal) and abdominal (ventral).*

### 3.3 Scan electron microscopy observations

The shape of the python, as shown in figure 1-a, is essentially non uniform. The perimeter of the cross sections of the along the body are not equal. In addition the build of the snake is essentially stocky. As a consequence, the mass of the animal will mainly be concentrated in the middle section of the trunk and the vicinity. Such a form reflects on the sliding behavior of the beast and on the reaction forces that the snake will experience during sliding. To this end, we can consider that the bulk of the friction-induced tractions where affect the region where most of the body mass is concentrated.

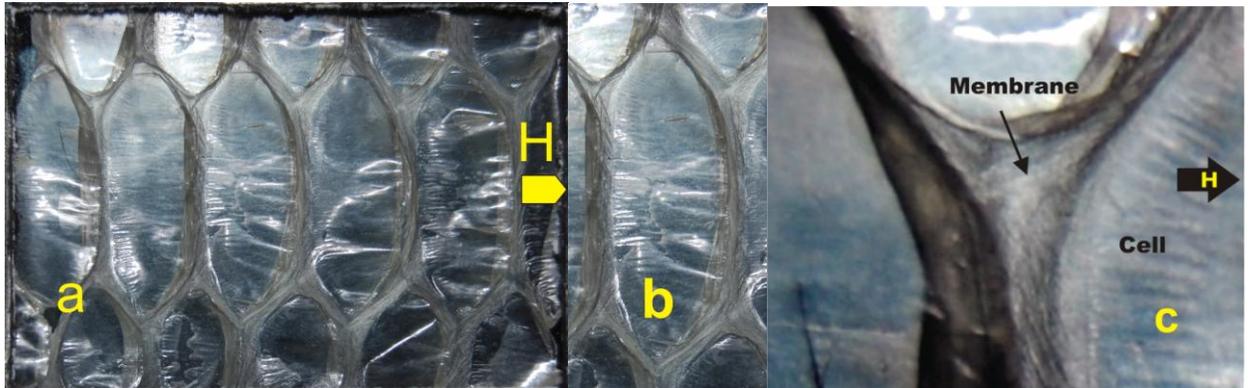

*Figure 6: The details of dorsal scales from the inside of shed skin. The terminology used is: membrane to denote the major area of the scale and boundary to denote the raised part forming the circumference of the scale.*

A consequence of such a preposition is that the extremes of the trunk region, namely the neck and the tail regions, will exhibit different friction loading and will undergo different sliding contact mechanics than the bulk of the trunk region. This, in turn, requires, from a functional design point of view, that the local surface geometry and topographical features within the two extremes of the trunk be distinctly different from those of the mid-section. In this sense, an analogy between the body of the snake and the surface geometry of a cylinder bore may be developed. Such an attempt draws upon the positions that a piston assumes while sliding during a combustion cycle of an ICE.

In an ICE a piston assumes three positions. These are the Top Dead Position (TDP), the Bottom Dead Position (BDP), and the Mean Position (MP). Each of these positions has a distinct lubrication requirement due to the manner that the piston engages the cylinder walls. Further, the local lubrication mechanism operating in each position is different. This, in turn, requires that the local surface texture be compatible with the lubrication regime necessary to maintain proper function and minimize damage. Similar to a snake the cylinder requires different textural features at the extremes, top and bottom, of the sliding stroke. An analogy is, thus, drawn between the main section of the trunk and the MP. Similarly, the TDP and the TDP are associated to the neck-trunk and the trunk-tail regions respectively.

To evaluate the parameters of the skin we identified 12 (twelve) points on the reptile hide. These are shown in figure 7. The general location of the selected skin samples are labeled by roman numerals. Table 1 presents a summary of the non-dimensional position of each area chosen for analysis. The values expressed as x/l entries were obtained by referring the actual distance between the centroid of the respective ventral scale and the nose tip of the reptile to the total length of the snake. Thus the x/l value represents the distance from the nose point, N, to the centroid of the particular ventral scale divided by the distance from the nose to the tail point T which is roughly about 11270mm. The roman numerals in the figure denote key locations on the hide from which skin swatches were selected for SEM observations. The points denoted I. and IV identify ventral scales located at the top and bottom boundaries of the trunk section. Thus, point I identifies a ventral scale roughly located within the TDP region and point IV identifies a scale located within the BDP region. Points II and III meanwhile refer to scales located on the front half of the skin and the rear half of the skin respectively. Both scales, however, are located within the MP region.

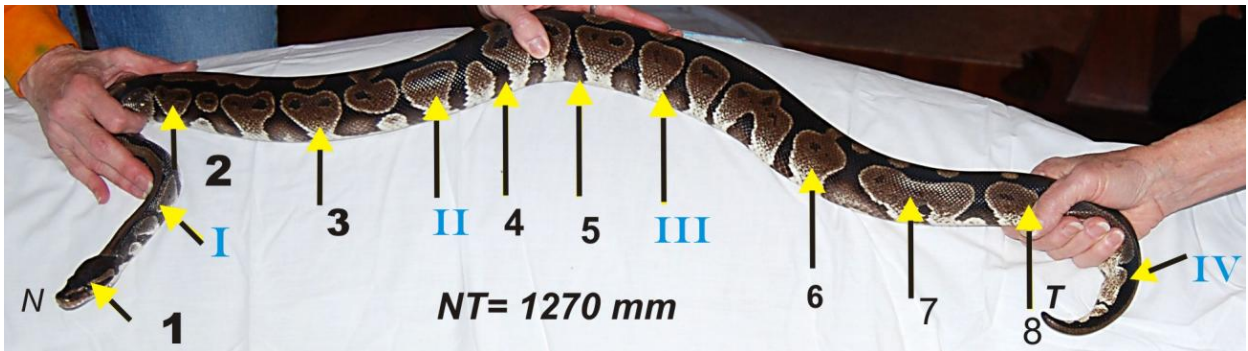

*Figure 7 Positions chosen on the shed skin of snake for observation and characterization of dimensional metrology. Roman Numerals indicate principal load bearing transition regions on the hide of the snake*

*Table 1: Relative position of skin samples chosen for observation and surface parameter evaluation.*

| Position | 1 | I | 2 | 3 | II | 4 | 5 | III | 6 | 7 | 8 | IV |
|---|---|---|---|---|---|---|---|---|---|---|---|---|
| X/L | 0.01 | 0.15 | 0.25 | 0.45 | 0.55 | 0.6 | 0.65 | 0.75 | 0.8 | 0.85 | 0.9 | 0.975 |

Skin swatches from each of the chosen positions were examined at different magnifications (X=250-X=15000) in topography mode. In order to suppress charging phenomena and improve the quality of observation, the surface of each sample was coated with a 10nm thick layer of platinum (Pt) using a sputter coater (EMITECH K575X).

For each position, samples from the dark and the light colored skin (see figure 5) were also examined along with samples from the underside of the body. Major features of the observations are shown in figures 8 and 9.

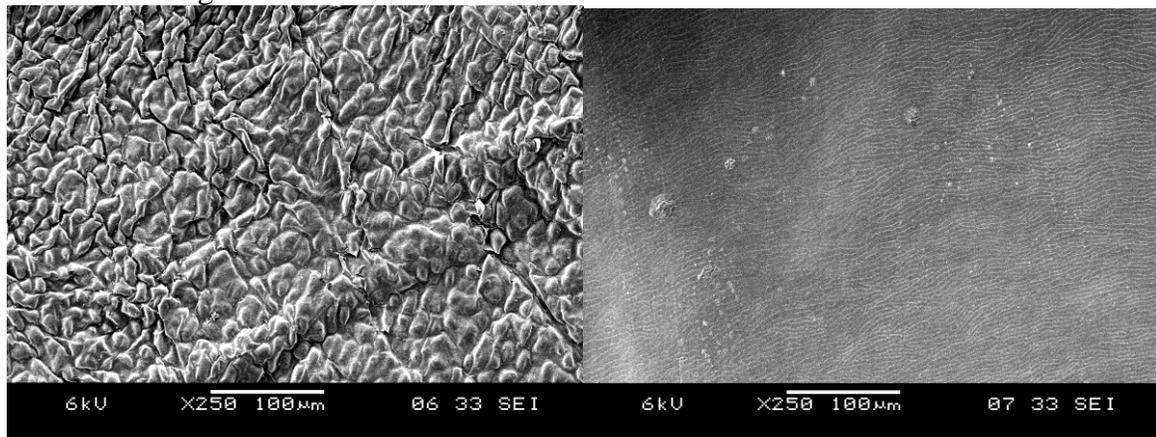

a-ventral scales- Scale boundary  b-ventral scales: Membrane

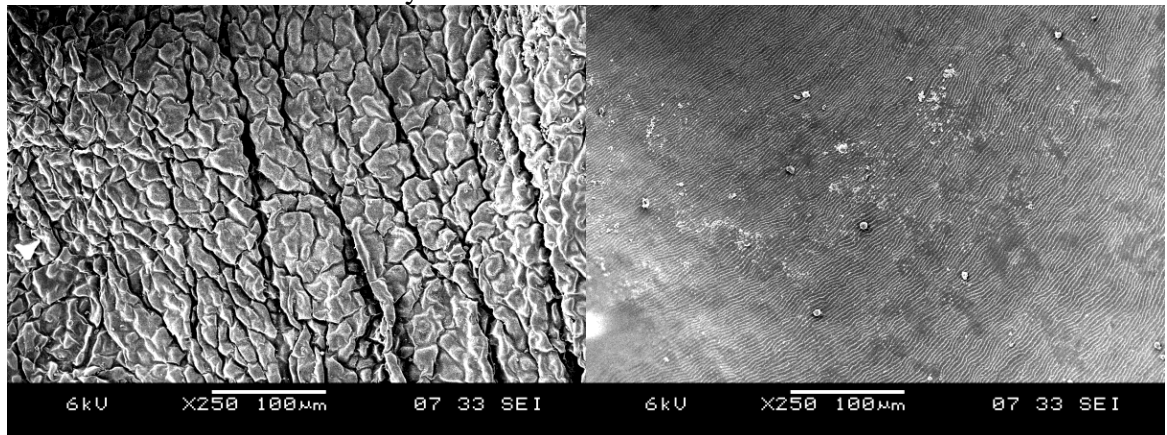

c-dorsal scales- light skin boundary  d-dorsal scales- light skin Membrane

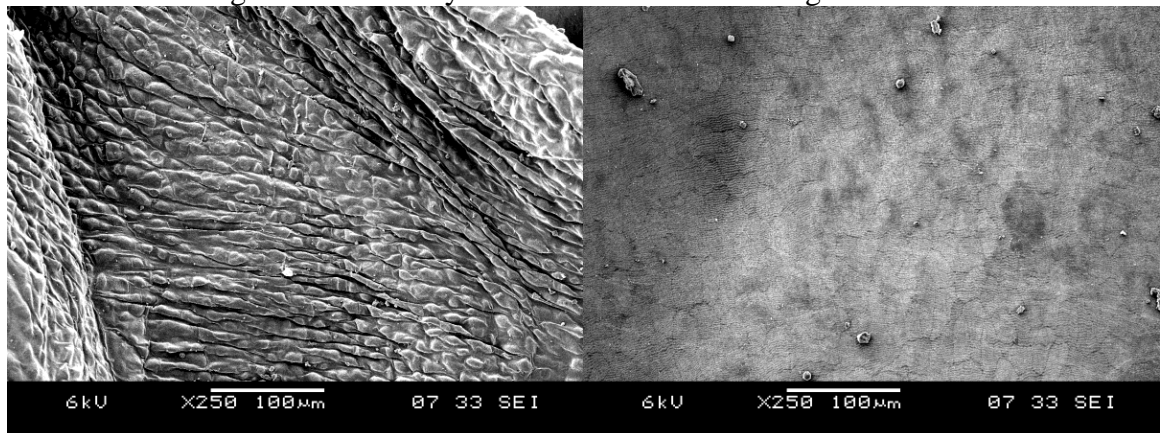

e-dorsal scales- dark skin boundary  f-dorsal scales- dark skin Membrane

*Figure 8: Major features of SEM observations of the skin swatches. Magnification used is X-=250 and the scale bar is 100 µm.*

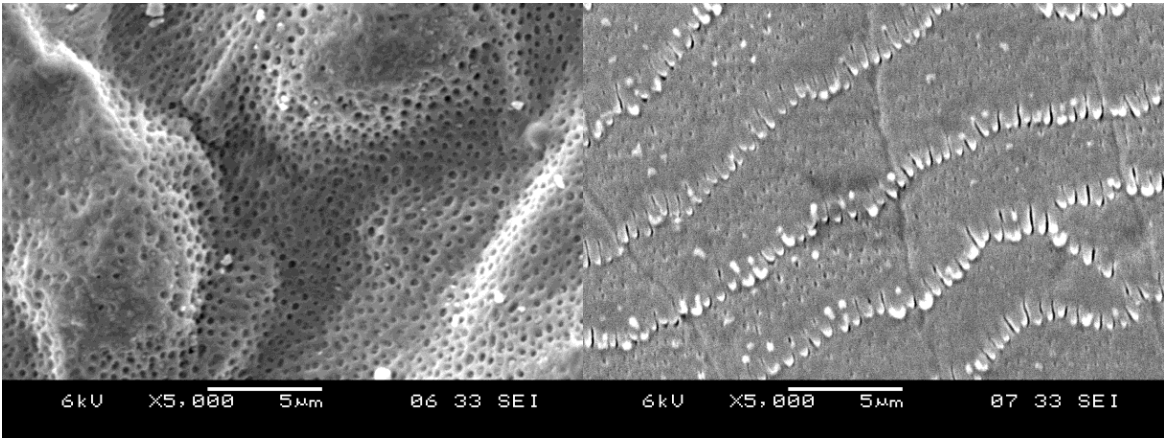
a-ventral scales- Scale boundary      b-ventral scales: Membrane

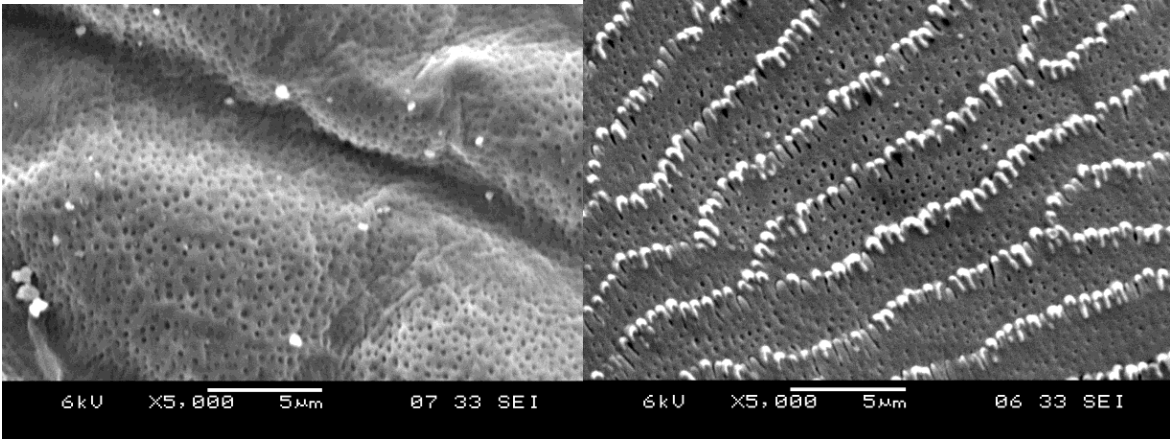
c-dorsal scales- light skin boundary      d-dorsal scales- light skin Membrane

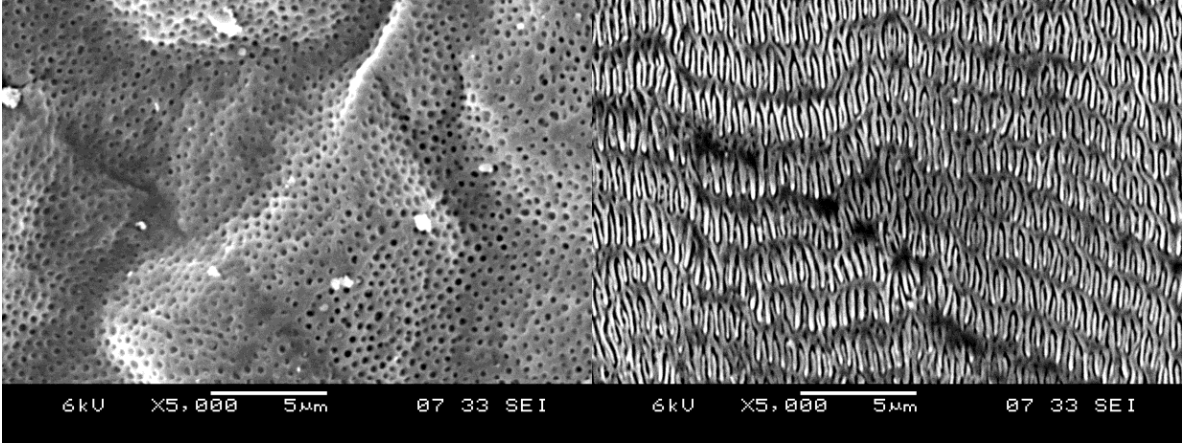
e-dorsal scales- dark skin boundary      f-dorsal scales- dark skin Membrane

*Figure 9: Major features of SEM observations of the skin swatches at a magnification of X=5000. The scale marker is 5 μm. Pictures are in one-to-one correspondence with those provided in figure 10.*

Figure 8 (a-e) depicts SEM micrographs taken at a magnification of X=250. Whereas, figure 9 (a-e) provides the micrographs of the same skin swatches depicted in figure 8 at a magnification of X=250. Thus, the pictures provided in figure 9 are at one-to-one correspondence with those provided in figure 8. The scale marker in figure 8 is 100μm and that in figure 9 is 5 μm.

Photographs of two positions are depicted in the figures. In both figures, the left hand side columns (labeled a, c, and e) depict photographs of a zone at located within the center of the boundary of the cell (i.e., within the elastic connector tissue between two cells). The pictures labeled (b, d, and f), within both figures, detail zones within the scale membrane. From top to bottom, within each figure, the photographs labeled a and b depict details of the ventral scales, those labeled c and d depict scales within the bright (light) dorsal skin, and finally those labeled e and f provide details of scales within the dark skin.

Figure 8 reveals a grainy appearance of the scale boundary irrespective of the side of the body (dorsal or ventral). The membrane structure, on the other hand, reveals a wavy appearance at the low magnification. This grainy appearance, within the scale boundary, manifests micro protrusions that appear to be of random shape and distribution. Size and volume of these protrusions appear to be quasi-uniform. The protrusions at this scale appear to be manifest folds within the elastic tissue. The wavy appearance of the membrane, however, appears to be in an overlapping arrangement. The spacing between waves seems to be more compact on the dorsal scales (figures 8 d and f) than the spacing within the ventral scales (figure 8-b).

Figure 9 reveal that the protrusions within the scale boundaries are approximately hemispherical (or at least concave clusters) and comprise pores. The space between the waves in the ventral scales, as well as the located dorsal scales is also full of pores. Dorsal scales located within the dark skin don't appear to contain pores. Two types of pores (or micro pits) may be distinguished: those located within the boundary and those located within the membrane. Image analysis of the pictures indicates that the diameter of the boundary-pores ranges between (200 nm – 250 nm). The diameter of the membrane-pores was estimated by Hazel et al [16] using AFM analysis to be in the range of 50nm to75 nm.

The surface of the membrane also comprises micro-nano fibrile structures. These are not of consistent shape and spacing. Note for example that the shape of fibril located in the dark colored skin region is different than that located within the light colored skin region (compare the X-5000 pictures). The fibrils in the dark colored scales are longer than those in the dorsal bright skin and those in the ventral scales. The width of the fibrils, in both regions, appears to be different. Fibril tips point toward the tail. Within the dark dorsal scales, fibrils are tapered and have a sharp tip. Scales within the bright colored and the ventral regions have a more rounded tip and appear to be of uniform width throughout the fibril length. Moreover the density of the fibrils seems to be different within the different color regions (denser within the dark colored region).

For each point shown in figure 7, a series of five SEM pictures at different locations within the particular ventral scale were recorded. The pictures were further analyzed to obtain fibril geometric information (counts, distance between fibrils, and length of individual fibrils). The average (arithmetic mean) of the information sets of the selected ventral scales were then plotted against the non-dimensional distance. Here we present the variation in the distance between waves of fibrils. To verify this observation we measured the distance between rows of fibrils (wave spacing) from SEM micrographs taken at each of the positions depicted in figure 7. In all, twelve positions were examined.

Figure 10 presents a plot of the internal spacing, $\lambda$, (distance between fibril rows in µm) as a function of the non.-dimensional distance X/L. Location on the skin may be obtained by comparing the X/L values to entries in table-1. Internal spacings in the figure represent the average of five separate measurements within different regions of the same SEM picture. As such, each data point in the figure is actually an average of 25 readings on the individual ventral scale. Approximate boundaries of the trunk are located within the shaded rectangle in the plot.

Data from the figure indicate that the distance between fibril waves vary between 3.5μm <l <.4.8μm in the ventral scales. The variation of that distance within the dorsal scales is 2 μm < λ< 3 μm and 1.5 μm < λ< 2.4 μm for the light colored and dark colored dorsal skin respectively. The shorter spacing is roughly located within the non-load bearing portions of the body (i.e.; the head and tail sections). The distribution of the separation distance λ along the body is not uniform. The internal spacing is larger within the general region of the trunk. The maximum spacing is roughly located within the middle section of the trunk (MP-S). It to be remembered that the arrangement of-the fibrils don't constitute straight lines. Note that the internal spacing between different rows of fibrils also differs by skin color and position within the body in the order $\lambda_{ventral} > \lambda_{dorsal\ light\ skin} > \lambda_{dorsal\ dark\ skin}$.

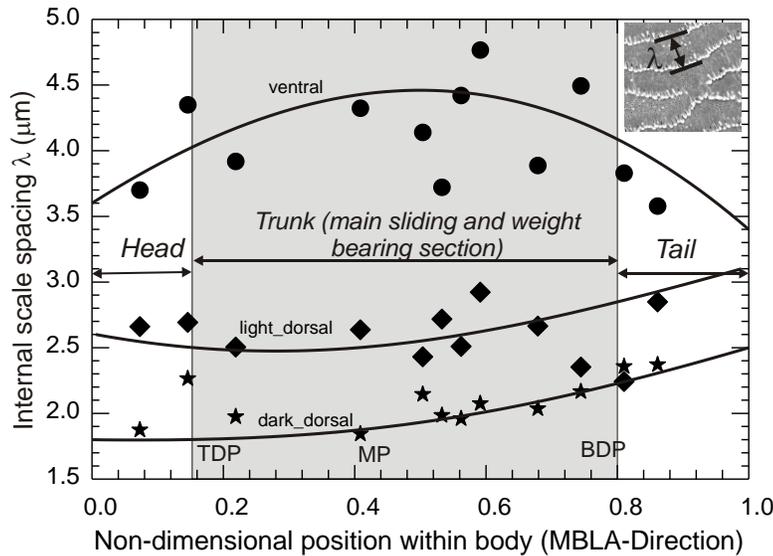

*Figure 10: variation in the intra-spacing between rows of the micro fibrils located within the membrane (hinge) of the snake scales as a function of: distance along the MBLA, position of scales and color.*

Further analysis of images revealed that the density of the boundary-pores vary by position. The number of pores per unit area is not constant along the body. It changes relative to the position within the skin. Figure 11 is a plot of the variation in the density of the pores relative to the two sides of the skin (back-Dorsal scales and abdominal-Ventral scales) and in relation to the color of the skin (Light Patches Vs Dark Patches) within the back also.

Density of pores was obtained from SEM micrographs of the by counting the pores in the picture and dividing by the area of the region scanned by the picture. Each point in the graph is an average of five counts taken from five different SEM observations of regions located within the same general area of analysis. Consistent with the trend noted in figure 10, the plot indicates that the pore density is higher in the order Ventral> Dorsal light > Dorsal dark.

4. **Metrology of the surface**
   **4.1 Topographical metrology**

To characterize the surface topography of the skin, we selected several swatches of skin (1500 μm by 1500 μm) for examination using White Light Interferometry (WLI). The results yielded the basic parameters that describe the surface (asperity radii, curvature etc.). Figure 12 depicts a typical WLI graph of the skin. The shown inteferogram pertains to a skin spot that is located along the waist of the snake from the belly side (ventral). Two interferograms are depicted: the one to the right hand side of the figure represents the topography of the cell-membrane whereas the one depicted to the left represents a multi-scale scan for the whole skin swatch. Note the scale on the right of the pictures as it indicates the deepest valley and highest point of the skin topography. For these skin swatches, the value of the deepest part of the membrane was about 120 μm, whereas the highest summit is about 100 μm. The comparable values for the whole swatch are about 5.5 μm and 8.2 μm respectively. These initial observations prompted the study of the so called Abbott-Firestone load bearing curve at different locations within the skin.

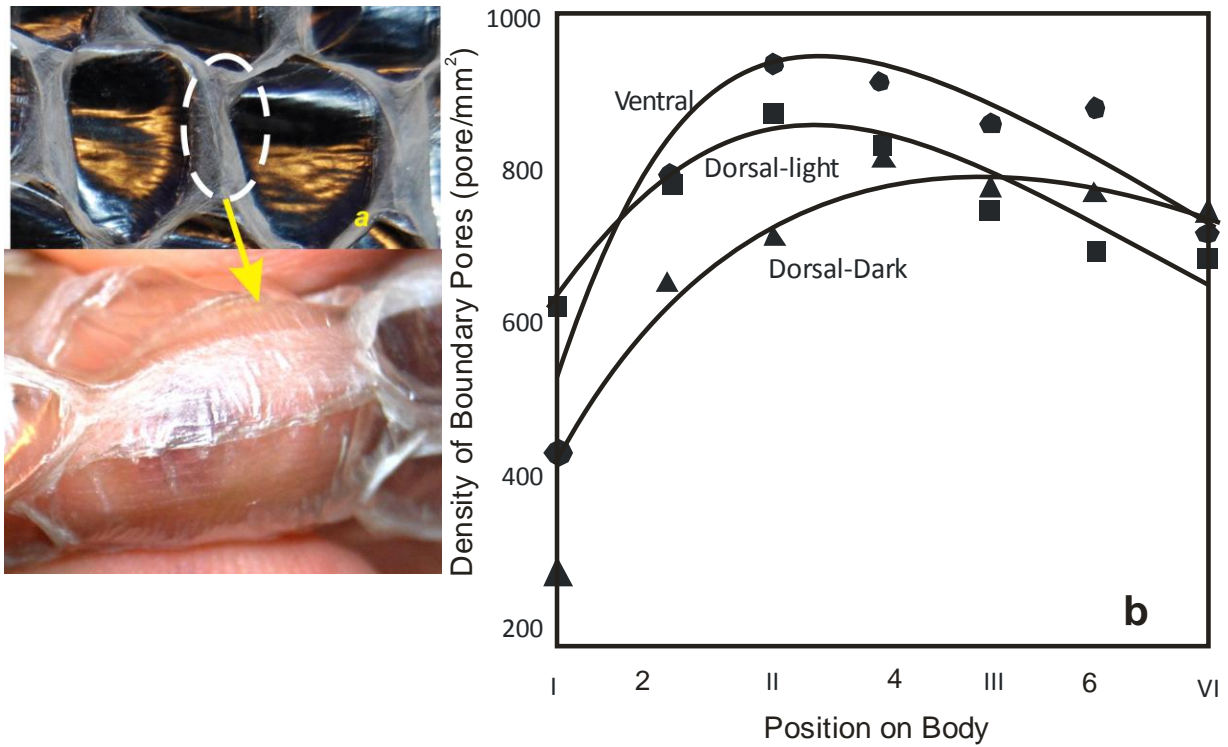

Figure 11. *The variation in the density of the boundary pores (pore /mm$^2$) with position, and with color of skin.*

### 4.2 Bearing curve analysis

Surfaces, irrespective of their method of formation, contain irregularities or deviations from a prescribed geometrical form. The high points on the surfaces are referred-to as asperities, peaks, summits, or hills and the low points as valleys. When two rough surfaces acted upon by a normal force come into contact, the opposite surface peaks to make contact first, are those for which the sum of the heights is the longest. As the load is increased, new pairs of opposite peaks having an even smaller sum of heights will be coming in contact. Once in contact, the surface peaks become deformed. This deformation leads to an increase in the contour area of contact, and as a result, to an increase in the number of peaks sustaining the load. Since the peaks differ in height, the deformation of various peaks on one hand the same surface will be different at any instant of time. The irregularities of the mating surfaces are out of contact over a considerable portion of the apparent contact area because of surface waviness and form errors. Whence, only of a fraction of the apparent area establishing contact will actually bear a load. A question that arises in such a situation is: given a known surface roughness profile, how can we calculate the area that actually supports a load? The answer is normally formulated in terms of a so called Load Bearing Area Curve (also known as the *Abbott-Firestone Load Curve* (AFLC)). The idea of that curve is to calculate the probability of a roughness protrusion to of a given height establishes true contact with a virtually smooth surface. Computing such a probability for a series of heights that sufficiently describe the surface, from a statistical point of view, yields a probability density distribution that relates to the true profile of the surface. Upon integrating this distribution with respect to surface height, we obtain the AFLC [40-42]. Study of the AFLC yields a worthy prospective of the potential behavior of a given surface upon sliding and the potential for damage through wear. Figure 13 presents the AFLC for three ventral scales on the live snake. These are highlighted in the photographs labeled 1, 2, and 3 in the figure. The picture labeled 1, depicts a ventral scale located in the middle section of the trunk (MP region). The picture labeled 2, depicts a ventral scale located within the general area of the throat-neck (TDP region). Finally, the photograph labeled 3, depicts a ventral scale located within the tail section. Plots in the figure are labeled accordingly.

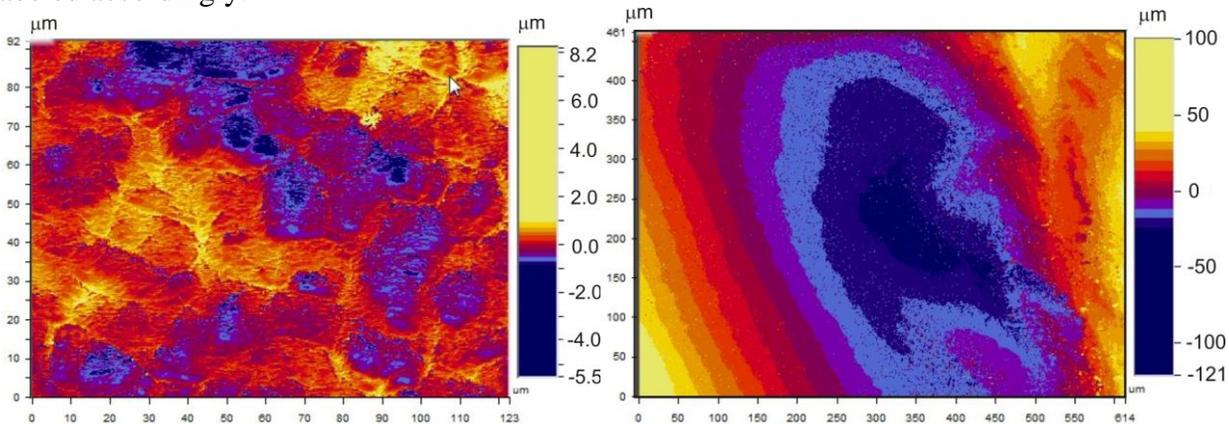

*Figure 12. Multi scale WLI graphs depicting the topography of the skin building block (Scale) boundary and membrane.*

The plot reflects the symmetry of the ventral scales. The individual plots are rotationally symmetric around the vertical axis (percentage data cut). On the other hand, there is a significant difference in the load bearing capacity of the middle section, position 1, and the other two regions (2 and 3). This is interesting as it supports the customization in natural design. As mentioned earlier, position 1 represents the zone where the bulk of the mass of the snake is contained. It also represents the region, within the ventral side of the snake, where most of the frictional tractions generated while locomotion is likely to be concentrated. In contrast to this region zones 2 and 3 are not likely to be as loaded. Consequently, the probability of sustaining severe damage in sliding is not prominent. From a design point of view, there is no requirement enforce the thickness of the surface. It is also apparent from the figure that different zones within the skin have variable surface profile parameters that are specific to their frictional profile.

Further analysis, the AFLC yields predictive information about sliding performance. The predictions formulate standardized surface-functionality assessment parameters. Of interest in this presentation is the so-called $R_k$ family of parameters [44-45] and their equivalent CNOMO counterparts [46].

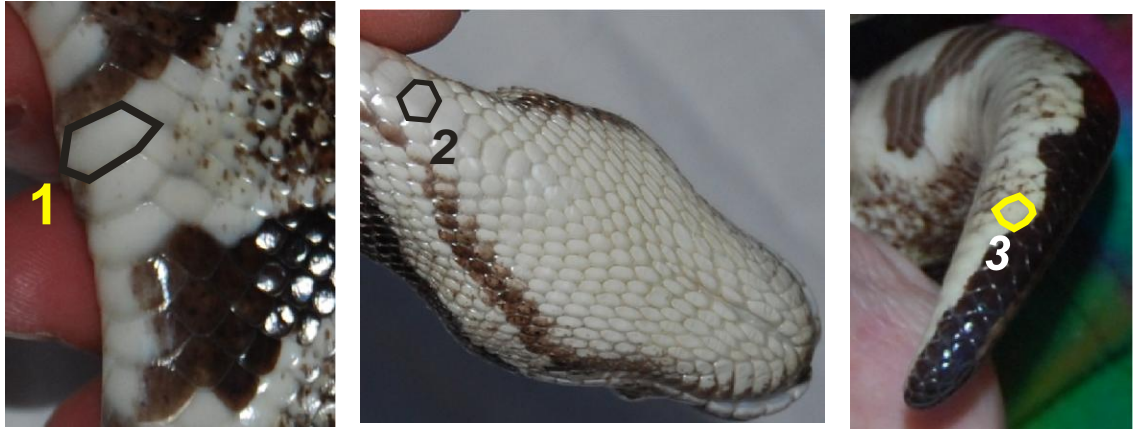

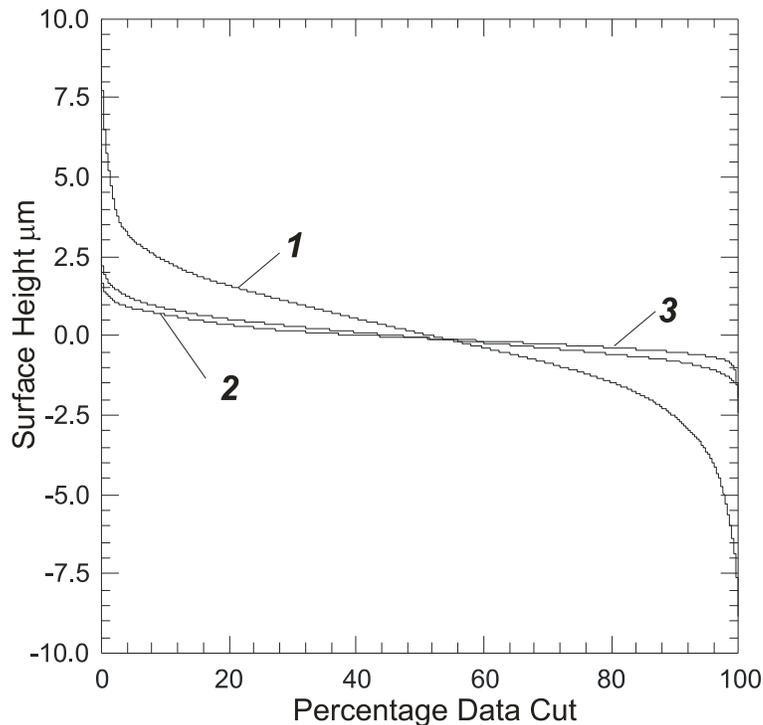

*Figure 13 Plot of the Abbott-Firestone Load Bearing Curve for three ventral scales, 1-middle section of the trunk, 2-neck-trunk boundary, and 3-tail section.*

We studied the load bearing characteristics of the skin at each of the key positions (I through IV). Surface parameters were extracted from SEM topography photographs. The complete set of analyzed pictures provided a matrix of roughness parameters that describe the texture of the shed skin at variable scales ranging from X-100 to X-5000. Table 3 (a and b) provides a summary of the parameters extracted from the analysis. It can be seen that the scale of the analysis affects the value of the parameters, which may point at a fractal nature of the surface.

Comparing the ratios between the Reduced Peak Height *Rpk,* Core Roughness Depth *Rk,* and Reduced Valley Depth *Rvk* reveals symmetry between the positions (compare the columns *Rpk/Rk*, *Rvk/Rk*, and *Rvk/Rpk* of table 2-b, and figure 14 a and b). This symmetry is interesting on the count that positions II and III represent the boundaries of the main load bearing regions (trunk). This is the region within the body where the snake has most of its' body weight concentrated and is the region that principally drives locomotion. The symmetry in surface design ratios is more apparent at higher magnification (X-5000). This implies that the symmetry is more significant at small scale, which points out at the uniformity of the surface basic building blocks at smaller sizes. Such symmetry may relate to the wear resistance ability of the surface or to the boundary lubrication quality of locomotion. Of interest also, is to find if implementing a surface of such characteristic parameters (functionally textured surface) in plateau honing for example would be conducive to an anti-scuffing and economical lubricant consumption performance.

*Table-2 effect of magnification on surface parameters as deduced from SEM-Micrographs*
a-   Surface parameters based on X-250 pictures

|              | Cr/Cf | Cl/Cf | Rpk/Rk | Rvk/Rk | Rvk/Rpk |
|---|---|---|---|---|---|
| Position I   | 0.718 | 0.861 | 0.391  | 0.159  | 1.144   |
| Position II  | 2.011 | 2.010 | 0.612  | 0.545  | 0.656   |
| Position III | 2.066 | 1.628 | 0.733  | 0.436  | 0.621   |
| Position VI  | 1.388 | 0.926 | 0.617  | 0.195  | 0.749   |

b-   Surface parameters based on X-5000 pictures

|              | Cr/Cf | Cl/Cf | Rpk/Rk | Rvk/Rk | Rvk/Rpk |
|---|---|---|---|---|---|
| Position I   | 1.930 | 1.207 | 0.654  | 0.285  | 0.679   |
| Position II  | 1.273 | 1.803 | 0.478  | 0.404  | 0.812   |
| Position III | 1.671 | 1.622 | 0.484  | 0.359  | 0.800   |
| Position VI  | 1.772 | 1.158 | 0.636  | 0.260  | 0.688   |

The motivation for such a proposal stems from identification of the common, functional and geometrical, features between the surface of the python and that of a honed surface.

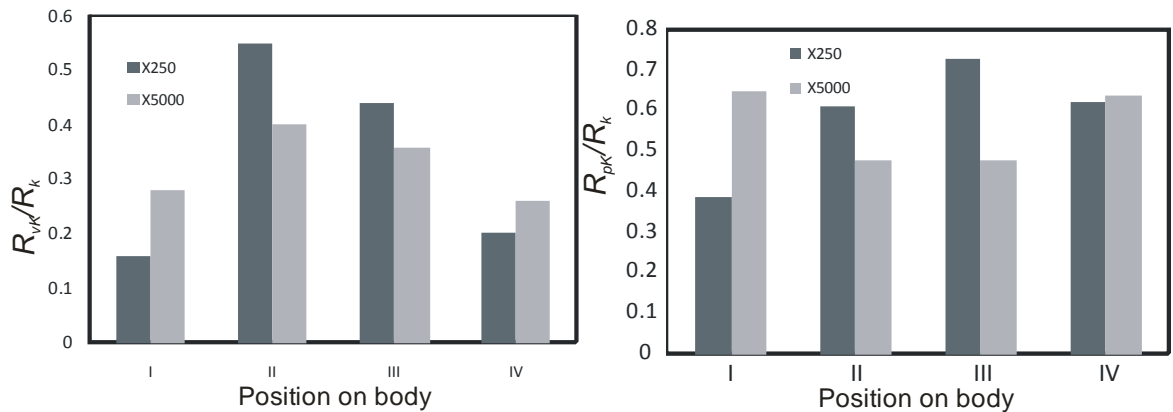

*Figure 14: Plot of the ratio of the load bearing parameters Rvk/Rk and Rpk/Rk at two magnifications X-250 and X-5000.*

## 5. Correlation to honed surfaces

One of the major requirements for an optimal honed surface is connectedness. Through perfect connectedness between all the surface unit-texture features, high lubrication quality, and economical lubricant consumption, is supposed to take place. Economical oil consumption, in essence, takes place because of controlled bearing-curve surface features (e.g, Rpk and Rvk). To date, there are no standardized values of relevant surface parameters to ensure superior performance of a honed cylinder liner. Instead, each manufacturer has in-house set of ranges that surface parameters are supposed to fall within for quality performance. To this end, a comparison between the basic metrological features of a Python skin and those of a, medium quality surface (say) would highlight, at least qualitatively, an optimal range of the geometrical proportions that should be maintained within a high quality performing surface.

To compare skin and honed surface, we examined segments of a commercial unused engine prepared by a sequence of honing processes. Figure 15 (a and b) depicts two SEM micrographs of one of the segments used for comparison with skin at two magnifications (X-250 and X-5000). All pictures were obtained after cleaning the surfaces by acetone in an ultrasonic bath. To obtain surface geometry parameters each SEM-micrograph was analyzed using the same program used to analyze the skin samples.

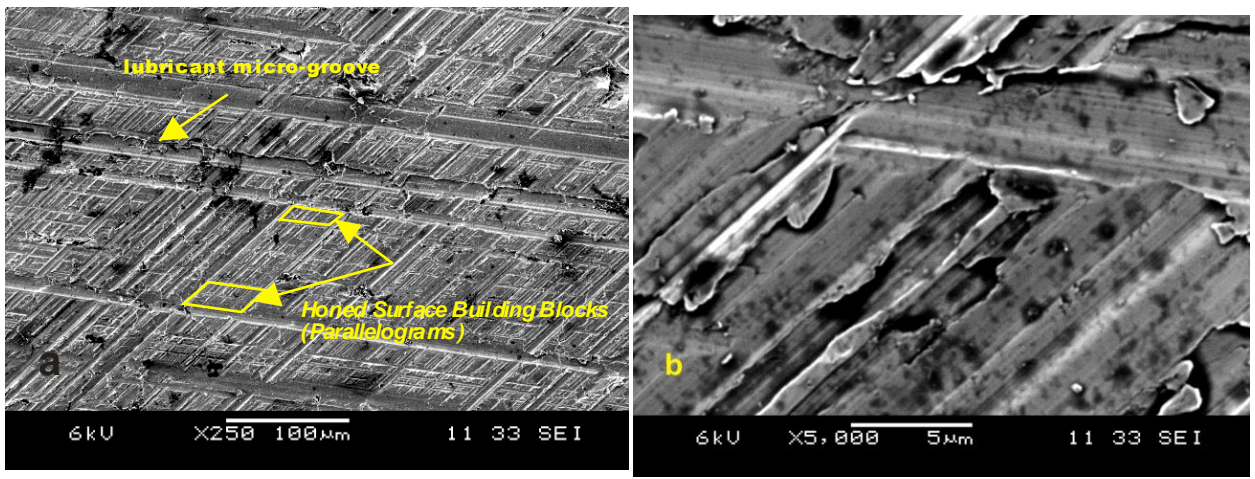

*Figure 15: SEM-Micrographs of a cylinder liner sample used for comparison with the Python skin, X-250 left hand side picture and X-5000 right hand side picture.*

Results of the analysis are given in figure 16: a and b compared to those extracted from the skin of the Python. The figure depicts metrological surface parameters calculated at two scales of observation: X-250 and X-5000. The parameters designated Python were obtained by averaging the values of the particular ratio of the zones II and III on the python skin. The plateau-honed surface (designated as P.H in figure) is classified as a so-called grade three surface [47]. According to the standard adopted in this case such a surface while not a superior finished surface is still acceptable from a quality control point of view. As such the comparison made in figure 16 is between the geometrical surface proportions of the main sliding zone in the Python (zones II and III) and the parameters of an average quality surface that is applied in real practice.

To establish a *qualitative* reference for analysis we recall the relation of each of the surface parameters $R_{pk}$, $R_{vk}$, and $R_k$ to surface performance in sliding. $R_{pk}$ is related to the amount of the surface that will be worn away during the run-in period, which is preferred to be at a minimum. $R_k$ relates to the working portion of the surface that will carry the load after the run-in period, which is preferred to have a relatively higher value. $R_{vk}$ meanwhile relates to the lowest portion of the surface that will retain lubricant, which for minimum oil consumption is required to be relatively high. As such, again qualitatively, the ratio $R_{pk}/R_k$ should be in the neighborhood of 0.5 or less, $R_{vk}/R_k$ should be relatively smaller than unity at or around 0.5, and $R_{vk}/R_{pk}$ should be relatively high at or around unity or slightly higher. Again we emphasize that such limits are qualitative estimates for the sake of comparison. Thus, proceedings from these estimates one would notice that the Python skin is slightly off the target limits at large scale (X-250). An opposite trend however is noted on smaller scale (X-5000). Interestingly, however, while the ratio $R_{pk}/R_k$ is almost within qualitative limits for the honed surface at large scale (X-250), it significantly departs from that limit at smaller scale (X-5000). The opposite is noted for the ratio $R_{vk}/R_{pk}$. Such an observation prompted the calculation of the percentage of change in the three examined geometrical proportions as a function of scale of observation. Results are plotted as figure 17.

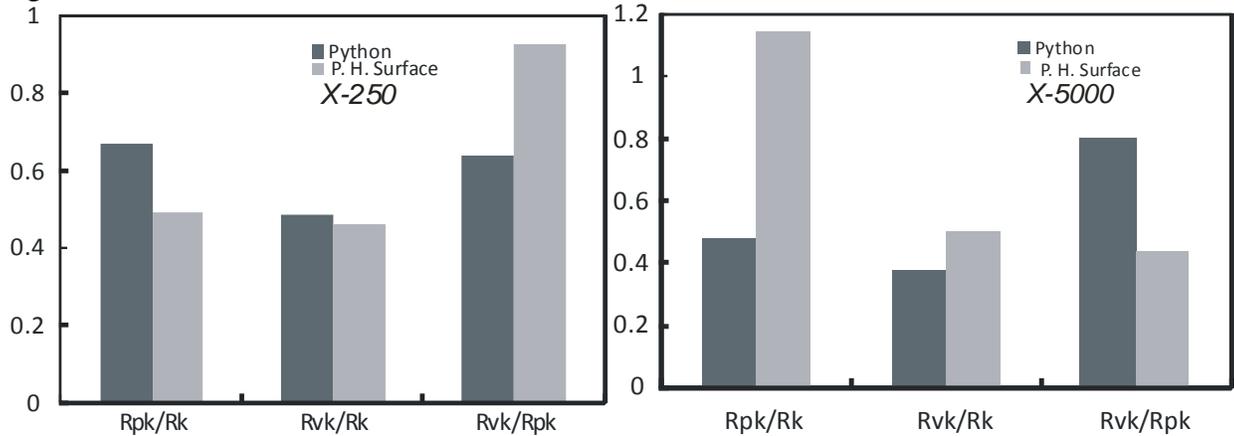

*Figure 16: Comparison between the geometrical and metrological proportions of the Python skin surface and those of a Plateau honed cylinder liner.*

Observe the Python skin has uniform and minimal, variation in the surface geometrical proportions than the honed surface. For the honed surface, the ratio $R_{pk}/R_k$ shows the largest variation. Such a wide variation may affect performance of the honed surface at very small scale through affecting the connectedness of the microgrooves and thereby the unobstructed flow of the lubricant while sliding. observation scale as seen in figure 17. In all there are many similarities between the honed surfaces and the geometry of Pythons, and snakes in general, that warrant additional studies.

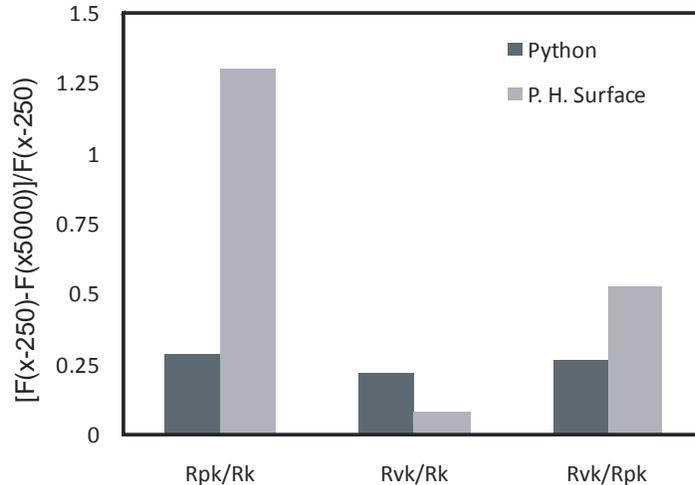

*Figure 17: Absolute value of percentage variations in surface proportions as a function of scale of observation.*

The geometry of a honed surface, its texture and ornamentation, is introduced through the action of the honing tool. The pattern comprising the surface texture, similar to that of the skin of a python, is composed of small sized building blocks. In a honed surface the basic surface unit is a four sided polygon, a parallelogram, whereas in the python skin it is mainly a hexagon that differs in the aspect ratio according to location within the body. Similar to a python, the parallelograms in a honed surface protrude above the main surface. For optimal performance, a major requirement for honed surfaces is perfect connectedness of between all the grooves separating the unit building blocks of surface texture. Perfect connectedness ensures unobstructed oil flow for lubrication of sliding surfaces and retention of oil to replenish the surface with lubricant to ensure separation of surfaces upon contact. Connectedness is also ensured at a smaller scale which renders the Python skin more appropriate than that of the honed surface examined. An additional criterion for an optimal honed surface is the absence of wear debris in the grooves that are of the same or larger width than the groove. For pythons, and snakes in general, maintaining an unbreakable boundary lubrication film depends on preventing debris from the environment, or dust particles from clogging lubrication passages within the body. The mechanism responsible for preserving the integrity of lubrication paths in a snake depends on the consistency of both geometry and metrology of the skin surface. Such consistency is evident from the trend of change in proportions of the main surface parameters as a function of position

**6.    Conclusions and future Outlook**

In this work we presented the results of an initial study to probe the geometric features of the skin of the Python regius. It was found the structure of the unit cells is of regionally similar shape (octagonal and hexagonal).
Although almost identical in size and density, the skin constituents (pore density and essential size of the unit cell) vary by position on the body. Analysis of the surface roughness parameters implied a multi-scale dependency of the parameters. This may point at a fractal nature of the surface a proposition that needs future verification.

The analysis of bearing curve characteristics revealed symmetry between the front and back sections of the snake body. It also revealed that the trunk region is bounded by two cross-sections of identical bearing curve ratios. This has implications in design of textured surfaces that retain an unbreakable boundary lubrication quality and high wear resistance.
Clearly much work is needed to further probe the essential features of the surface geometry. Namely, the basic parametric make up of the topography and its relation to friction and wear resistance.